# Unveiling the charge density wave inhomogeneity and pseudogap state in 1$T$-TiSe$_2$

Kai-Wen Zhang[a], Chao-Long Yang[a], Bin Lei[b], Pengchao Lu[a], Xiang-Bing Li[a], Zhen-Yu Jia[a], Ye-Heng Song[a], Jian Sun[a,c], Xianhui Chen[b,c], Jian-Xin Li[a,c*], Shao-Chun Li[a,c*] a *National Laboratory of Solid State Microstructures, School of Physics, Nanjing University, Nanjing 210093, China.*

b *Hefei National Laboratory for Physical Science at Microscale and Department of Physics, University of Science and Technology of China, Hefei 230026, China.*

c *Collaborative Innovation Center of Advanced Microstructures, Nanjing University, Nanjing 210093, China.*

*Corresponding authors. Email addresses: jxli@nju.edu.cn (J.-X. Li); scli@nju.edu.cn (S.-C. Li)

**Abstract:**

By using scanning tunneling microscopy (STM) / spectroscopy (STS), we systematically characterize the electronic structure of lightly doped 1$T$-TiSe$_2$, and demonstrate the existence of the electronic inhomogeneity and the pseudogap state. It is found that the intercalation induced lattice distortion impacts the local band structure and reduce the size of the charge density wave (CDW) gap with the persisted 2×2 spatial modulation. On the other hand, the delocalized doping electrons promote the formation of pseudogap. Domination by either of the two effects results in the separation of two characteristic regions in real space, exhibiting rather different electronic structures. Further doping electrons to the surface confirms that the pseudogap may be the precursor for the superconducting gap. This study suggests that the competition of local lattice distortion and the delocalized doping effect contribute to the complicated relationship between charge density wave and superconductivity for intercalated 1$T$-TiSe$_2$.

**Key Words**：Charge density wave, Pseudogap, TiSe$_2$, Scanning tunneling microscopy, Scanning tunneling spectroscopy.

## 1. Introduction

1$T$-TiSe$_2$ has been widely studied in the past few decades particularly as one of the typical charge density wave materials in transition metal dichalcogenides (TMDs) [1-3]. It undergoes a 2×2×2 commensurate charge density wave (CDW) transition at ~200 K [2]. While the origin for the CDW transition still remains ambiguous [4-12], studies on 1$T$-TiSe$_2$ have been revived recently due to the superconductivity discovered in the Cu intercalated 1$T$-TiSe$_2$ [3]. The Cu$_x$TiSe$_2$ forms a dome shaped superconductivity phase beginning at $x$= ~0.04, and the highest transition temperature $T_C$ ~ 4.15 K is obtained at $x$ = ~0.08 [3]. In the following, it was also found that high pressure and electric gating applied to 1$T$-TiSe$_2$ can realize the superconductivity, giving a similar dome shaped phase diagram [13, 14]. Such a strong resemblance of the superconductivity phase diagram with high temperature superconducting (HTSC) cuprates, makes it desired to reveal the mechanism of TiSe$_2$ superconductivity.

Even though the transport measurements of Cu$_x$TiSe$_2$ exhibited that the CDW transition is sufficiently suppressed as the superconductivity state emerges [3, 15], photoemission studies gave controversial indications whether it is the competition or coincidence between these two collective electronic states [16, 17]. Furthermore, recent studies unveiled the emergence of incommensurate CDW phase above the superconducting dome, e.g., the CDW domain walls, which has been proposed to be connected with the superconductivity transition [14, 18-20]. The CDW modulation can be even detected in a region beyond the end of the superconducting dome [19].

Atomic scale studies have been performed focusing on the Cu intercalated 1$T$-TiSe$_2$ [20, 21], and have revealed the phase shifted boundary of 2×2 CDW domains and the enhanced local density of states (LDOS) at Fermi level, which is believed to favor the superconductivity transition. However, the impact of intercalation to the electronic structures of 1$T$-TiSe$_2$ have not so far been fully understood.

In this study, we focused on the spectroscopic evolution induced by both intrinsic and extrinsic defects. We firstly characterized the self-doped 1$T$-TiSe$_2$ by its native defects with atomically resolved STM/STS, and then tuned the electronic structure by doping potassium atoms on the 1$T$-TiSe$_2$ surface, and explored its evolution. We demonstrated the real-space inhomogeneity on the self-doped 1$T$-TiSe$_2$ surface, i.e., the coexistence of CDW 'normal' regions (with identifiable CDW gap) and "suppressed" regions (with reduced CDW gap). We found that a pseudogap emerges at the Fermi level, prior to the superconductivity transition, which is more prominent in the CDW "normal" regions. We attribute the decrease of CDW gap to the local lattice distortion induced by the intercalated atoms, and the formation of pseudogap to the doping effect by the delocalized donated electrons. Further electron doping via Alkali metal deposition leads to the formation of gaps with coherence peaks evolving from the pseudogap, suggesting that the pseudogap observed in this study may be a precursor to the superconducting gap.

## 2. Experimental methods

STM/STS characterizations were conducted in ultrahigh vacuum (UHV, the base pressure is 1 × 10$^{-10}$ Torr) with a low temperature STM (Unisoku Co. USM1600). The 1$T$-TiSe$_2$ single crystals were *in situ* cleaved in UHV, and then quickly transferred to the STM stage for scan. Alkali metal K was deposited with a flux of ~ 0.5 monolayers (MLs)/min. STM data were acquired at low temperatures varying from ~30 mK to ~30 K. STM images were taken with a constant current mode. The differential conductance d$I$/d$V$ spectra were taken via a lock-in amplifier with the ac modulation of 998 Hz.

## 3. Results and discussion

Fig. 1a shows the typical STM topographic image of a cleaved 1$T$-TiSe$_2$ surface, displaying a 2×2 charge density modulation. The native point defects, naturally introduced during the single crystal growth, are identified in Fig. S1 (online). Most of the defects can be assigned as the intercalated Ti atoms, and minor populations of Se substitution by iodine (I) / oxygen (O) or Se vacancies are also found, according to previous reports [22, 23]. The concentration of the Ti interstitial atoms is estimated as ~ 0.9%. Even though the surface exhibits a rather uniform 2×2 periodicity at relatively high bias energy, as shown in Fig. 1a with $U$ = +200 mV, the topography taken at lower bias, roughly near the edge of CDW gap, shows the noticeable inhomogeneity, see Fig. 1b with $U$ = –80 mV. In Fig. 1b, some areas of the surface exhibit a faded 2×2 electronic modulation, namely the region I, while the rest areas exhibits clear 2×2 modulation, namely the region II. The regions I and II exhibit an obvious difference between the CDW gap sizes in the d$I$/d$V$ spectra, see Fig. 1c. In region II, the CDW gap can be clearly identified as ~ 80 meV based on the obvious slope changes in the STS spectra, consistent with the values reported in previous studies [20, 21, 25]. However, in regions I the CDW gap size is drastically reduced. Regardless of the inhomogeneity of the CDW gap across the surface, the regions I and II exhibit indistinguishable 2×2 superstructure at higher bias energy, see Fig. 1a. Therefore, we name the region I and II as "suppressed" and "normal" CDW regions, according to the magnitude of CDW gap. The d$I$/d$V$ maps measured on both of the regions further confirm this electronic inhomogeneity, see Fig. 1d-g. At the bias voltage away from the CDW gap region, the d$I$/d$V$ map shows a continuous 2×2 superstructure with the fluctuation in the d$I$/d$V$ contrast, see Fig. 1d and 1g. However, at the bias voltage within or close to the CDW gap region, strong inhomogeneity is found in the electronic states the region I and II: only a small area holds the clear 2×2 superstructure but the rest of the region does not, see Fig. 1e and 1f. Such a strong inhomogeneity in the electronic states induces the strong variance in the STM images taken at different bias voltages.

d$I$/d$V$ curves taken along a line spanning over both the regions I and II disclose the detailed electronic evolution, as shown in Fig. 2a. As approaching from region I (the red dot in Fig. 2b) to region II (the purple dot in Fig. 2b), in addition to the spatial variation of CDW gap as discussed above, there also exists an interesting spectral weight transfer, particularly in the region II. Such spectral weight transfer leads to the accumulation of the density of states near the Fermi level, see Fig. 2a. Concomitantly, a V-shaped gap is always formed at $E_F$, referred to as a pseudogap. The depth of the pseudogap, from the peak to bottom, is more prominent in region II. Fig. 2c shows more clearly that the edges for the pseudogap locate at ~±10 mV. It is worthwhile noting that the formation of the V-shaped pseudogap is more preferred in region II than in region I. In contrast, there is no prominent pseudogap observed in the region I, see Fig. 2a. In the recent microscopic studies [20, 21], singularity or depression at the Fermi level, which is somewhat similar to our observation of the pseudogap, was also indicated in the data for Cu-doped TiSe$_2$ samples but with no detailed discussion.

At extremely low temperature of ~30 mK, a kink feature, located at ~±3 mV, is well resolved inside the V-shaped pseudogap, suggesting the emergence of another gap, as shown in Fig. 2c. We define this kink feature as the opening of a "small pseudogap" and thus two pseudogaps coexist in 1$T$-TiSe$_2$. The two pseudogaps are further distinguished by their different dependence on the applied magnetic field. Fig. 3 shows the evolution of the pseudogap upon the external magnetic field applied perpendicular to the TiSe$_2$ surface. The large pseudogap, as guided by the black arrows in Fig. 3a, can persist up to ~9 T under an applied magnetic field, while the "small pseudogap", as guided by the black lines in Fig. 3a and b, is smeared out at ~7 T, as shown in Fig. 3. Even though the origin that causes the pseudodap is still not clear, the magnetic field that smears out the pseudogap, i.e., ~7 T for the small one and at least > 9 T for the large one, is significantly larger than the critical field $H_c$ for superconducting phase in doped TiSe$_2$ [14], consistent with the studies for pseudogap in other materials [26-28].

Even though the origin of the pseudogap in 1$T$-TiSe$_2$ is not clear, we follow the studies of pseudogap in HTSC cuprates [29] and now investigate its temperature dependence. Fig. 4a shows the evolution of the d$I$/d$V$ curves taken at a specific surface location. It is indicated that the pseudogap can persist up to at least ~18.5 K, far above the optimal $T_C$ of ~4.15 K for Cu$_x$TiSe$_2$-based superconductivity [3, 13, 14] and below the transition temperature of CDW $T_{CDW}$ [2]. Based on the resistance-temperature curve, Fig. 4b, the $T_{CDW}$ is estimated as ~150 K, in agreement with previous reports for the pure or lightly doped 1$T$-TiSe$_2$ samples [2, 3, 24]. Furthermore, no coherence peaks are observed in the temperature range investigated. We thus rule out the possibility that the V-shaped gap is a superconducting gap.

To further investigate the evolution of the pseudogap upon electron doping, we deposit potassium (K) atoms that donate electrons to the 1$T$-TiSe$_2$ surface, see Fig. 5a-c. More STM images of K doped surface can be found in Fig. S2 (online). The K atoms are adsorbed on the surface in single atom format, and exhibit as the bright round protrusions. It is found that the magnitude of the V-shaped pseudogap systematically decreases with increasing K coverage, see Fig. 5d. Remarkably, as the K coverage reaches ~ 0.28 ML, the coherence peaks start to appear at the edges of the gap. According to the phase diagram of TiSe$_2$ [3, 19], it is reasonable to expect that these coherence-peaked gaps are related with the formation of Cooper pair. Such an assumption is supported by the experimental evidence that the coherence peaks can be suppressed under the magnetic field, see Fig. 5e. The BCS fitting of the STS data gives a typical superconducting gap of ~1.2–1.5 mV, larger than the previous estimation of 0.5 mV for the Cu doped TiSe$_2$ with $T_C$ of ~2.3 K [30], which may indicate a higher transition temperature for the K doped TiSe$_2$ surface. Such an electron-doping induced evolution indicates that the pseudogap in the lightly doped TiSe$_2$ may be the precursor to the superconducting gap. The evolution of the gap size as a function of K doping is plotted in the schematic phase diagram Fig. 5f.

The intercalated atoms usually introduce a few impacts to the host materials, such as the lattice distortion, and the donated carriers [31]. The former effect is expected to be more prominent in the proximity of the intercalated atoms, while the latter more delocalized over the surface. According to the surface locations of the defect atoms, as determined in the atomic resolution topographic images in Fig. 1a,b, it is not straightforward to connect the electronic inhomogeneity to the phase shifted boundaries of CDW domains, since the domain boundaries are one dimensional and directly run through the defective sites [20, 23], which is not the case for this observation. However, it is found that more defective atoms, e.g., Ti interstitials, are statistically hosted in region I than in region II, as is shown in Fig. S3 (online). Following the previous study on the Cu intercalated TiSe$_2$ [21], such inhomogeneous regions I and II can be connected to the local fluctuation of the defect concentration, mainly the Ti interstitials. The possibility that the localized non-dispersive defect states in the region I fill up the CDW gap is excluded, because the identified region I can be as far as a few nanometers away from the defects centers. Therefore, we attribute the decrease of CDW gap in region I to the band structure change induced by the local effect of intercalation.

The intercalation of Ti atoms induces the local lattice distortion and the bonding of the Ti intercalant to the neighboring Se lattice atoms. Both of the issues lower the hole-like Se 4$p$ band close to the $\varGamma$ point, according to the previous calculations [32-34] and thus reduce its band overlap with the Ti 3$d$ band around the $L$ point. Considering the electronic origin of CDW transition, for example, the electron-phonon or excitonic interactions, where the band overlap plays a key issue in driving the CDW transition, the reduced band overlap between Se 4$p$ band and Ti 3$d$ band will decrease the favorability of CDW transition, which would be indicated by the decrease of CDW gap. The other types of defects in TiSe$_2$, such as the substitution or vacancy of Se, can induce the band bending of the Ti 3$d$ band according to the previous study [22], which play a non-essential role due to their much less concentration.

We ascribe the formation of pseudogap in regions II as the result of free electron carriers contributed by the Ti interstitials or surface potassium. The Ti interstitial located in the van der Waals (vdW) gap of 1$T$-TiSe$_2$, and the potassium on the surface of 1$T$-TiSe$_2$, behave like an electron donor [22]: Shifting up the Fermi level towards the conduction band and thus enhancing the local density of state at Fermi level. Such an electron doping effect was also proposed to be a key factor to promote the superconductivity transition [20]. Prior to the instability to superconducting state, the pseudogap is formed upon doping electrons.

Different from the native defects in 1$T$-TiSe$_2$, the extrinsic electron donors, potassium (K) atoms deposited on the TiSe$_2$ surface, donate electrons to the TiSe$_2$ surface with negligible lattice distortion. Thus it is expected that the electron doping effect plays the predominantly role in the further evolution of the pseudogap into coherence-peaked gap.

From the microscopic view in this study, it seems not likely that the formation of pseudogap / superconducting gap compete with the local 2×2 CDW superstructure. Furthermore, the magnitude of CDW gap, which can be tuned via local lattice distortion, is not straightforwardly correlated with the spatial 2×2 modulation. For instance, even though the local lattice distortion drastically decrease the CDW gap, and break the long range CDW coherence, the local 2×2 domains can be still persisted. Such an argument can explain the controversial results observed recently with different experimental techniques [3, 15, 19, 20]. For example, the resistance measurements exhibiting the complete suppression of CDW phase upon the emergence of superconductivity, indicated the decrease of CDW gap [3, 15], while the XRD or spectroscopic measurements detected the local 2×2 superstructure within the whole doped region [19, 20].

**4. Summary**

In summary, we have characterized the lightly doped 1*T*-TiSe$_2$ with STM / STS, and observed the inhomogeneity of CDW gap and the emergence of pseudogap state. While the intercalation induced lattice distortion may change the local band structure, e.g., reducing the CDW gap, electron doping effect promotes the formation of pseudogap. These phenomena resemble to what have been commonly observed in HSTC cuprates and other Mott insulators. This study suggests that TMD materials may exhibit more complicated phenomena upon intercalation. Further studies are expected to be stimulated regarding to the low energy excitations and superconductivity.

**Conflict of interest**

The authors declare that they have no conflict of interest.

**ACKNOWLEDGMENTS**

This work was supported by the Ministry of Science and Technology of China (2014CB921103, 2013CB922103, 2016YFA0300400, 2015CB921202), the National Natural Science Foundation of China (11774149, 11374140, 11190023, 11774152, 51372112, 11574133), NSF Jiangsu Province (BK20150012), and the Open Research Fund Program of the State Key Laboratory of Low-Dimensional Quantum Physics.

**FIGURE CAPTIONS**

FIG. 1. CDW inhomogeneity on the 1T-TiSe$_2$ surface. (a) STM image of the cleaved 1T-TiSe$_2$ surface taken at ~4 K (bias voltage $U$ = +200 mV, tunneling current $I_t$ = 100 pA, size: 20 × 20 nm$^2$). (b) STM image of the cleaved 1T-TiSe$_2$ surface taken at ~4 K (bias voltage $U$ = –80 mV, tunneling current $I_t$ = –100 pA, size: 20 × 20 nm$^2$). The fast Fourier transform (FFT) of the STM images in (a) and (b) is shown in the corresponding insets. (c) Differential conductance d$I$/d$V$ spectra taken in the region I (blue) and II (red) at ~4 K, respectively. The corresponding CDW gaps are marked by the vertical lines. (d) to (g), d$I$/d$V$ conductance maps measured over the area (9 × 9 nm$^2$) as marked by the yellow dashed square in (a) and (b). The maps include both of the regions I and II. The applied bias voltages are shown in the maps. The red parallelogram marks the unit cell of 2×2 modulations. At low bias voltages near the Fermi energy, i.e., (e) and (f), the phase II area shows a prominent 2×2 electronic modulation and high intensity, but the phase I shows no clear modulation and low intensity. At high bias voltages, i.e., (d) and (g), both the phases I and II show indistinguishable 2×2 modulations, except for the intensity contrast.

FIG. 2. Spatial evolution of the CDW gap and pseudogap on the 1T-TiSe$_2$ surface. (a) Spatially resolved differential conductance d$I$/d$V$ spectra taken along the dotted line as marked in (b). The spectra are plotted in rainbow colored scale. From top to bottom, the spectra span from the region I to the region II. (b) STM images showing the topography where the d$I$/d$V$ spectra (a) were taken. The exact locations are marked by the dots with the same colors as (a). The up image (U = –80 mV) was taken close to the CDW gap, and the down image ($U$ = +200 mV) far away from the CDW gap. (c) A typical high resolution differential conductance d$I$/d$V$ curve taken in the region II at ~30 mK. The magnitude of pseudogap, $\Delta$, is estimated as ~10.6 meV, and the gap edge is marked by black arrows. The kink feature, forming a "small pseudogap" inside the V-shaped pseudogap, is located at ~±(3–4) mV and marked by green arrows.

FIG. 3. Magnetic field dependence of the pseudogap. (a) The evolution of d$I$/d$V$ spectra for the pseudogap taken on a specific lattice site under magnetic field at ~30 mK. The change of the pseudogap is not obvious, and under magnetic fields of 9 T, the pseudogap can be still well identified.

(b) The "small pseudogap" undergoes a clear evolution under magnetic field, and is gradually closed at ~7 T. The black arrows mark the positions of the pseudogap and the black lines the small pseudogap.

FIG. 4. Temperature dependence of the pseudogap and the resistivity. (a) Temperature evolution of the differential conductance d$I$/d$V$ spectra taken on a specific surface location. (b) $\rho$ vs $T$ curve measured from room temperature down to ~2 K.

FIG. 5. Dependence of the pseudogap on the potassium ($K$) doping. (a)–(c) Typical STM images ($U = -100$ mV, $I_t = -100$ pA; 30 × 30 nm$^2$) taken on the same surface with ~0.06 MLs, ~0.13 MLs, and ~0.28 MLs of potassium (K) atoms, respectively. (d) Spatially averaged d$I$/d$V$ spectra taken at the same surface of 1$T$-TiSe$_2$ with various coverages of K. The edges of the pseudogap are marked by black arrows, and the coherence peaks are highlighted in red for guiding eyes. (e) d$I$/d$V$ spectra taken on a specific lattice site under various magnetic field at ~30 mK, showing the evolution of the coherence-peaked gap under magnetic field. With increasing magnetic field, the coherence peaks are suppressed but the gap shape is kept. (f) Schematic phase diagram for K-doped 1$T$-TiSe$_2$ illustrating the relationship between the gap size $\Delta$ and the coverage of K doping. The red squares represent the experimental data of the gap size extracted from (c). The green colored area represents the CDW phase, the dark blue colored area the pseudogap phase (PG), and the yellow colored area the superconducting dome (SC) caused by K doping as determined by the gap with coherence peaks.

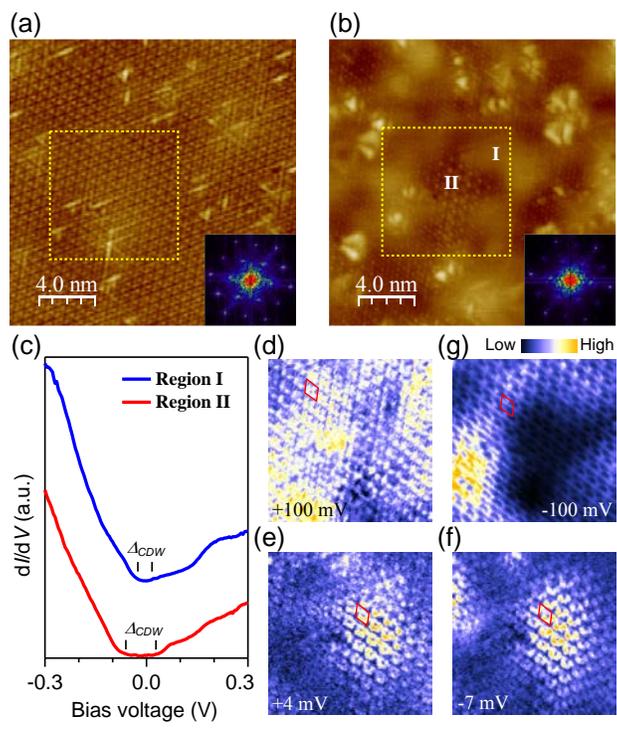

Figure 1

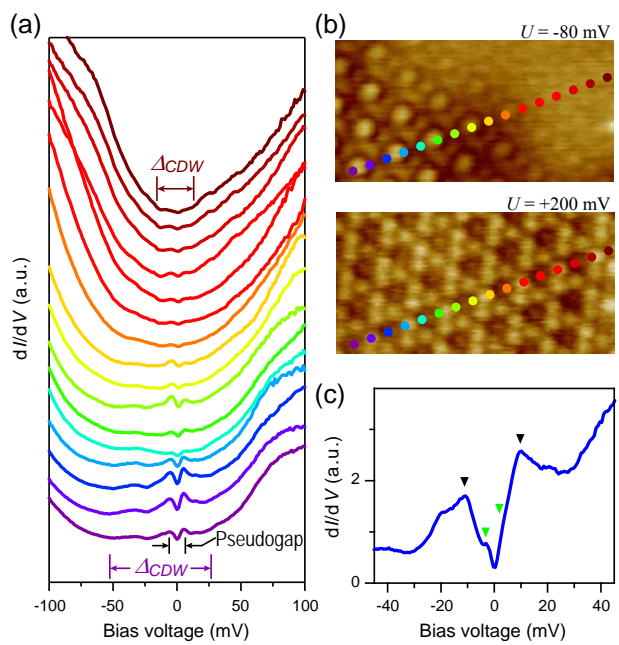

Figure 2

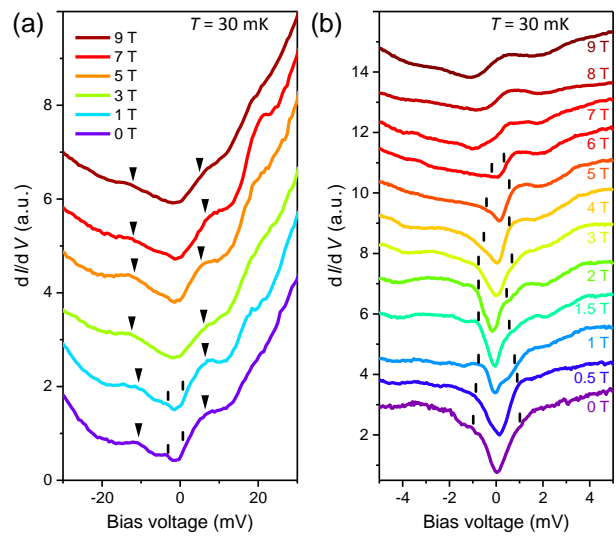

Figure 3

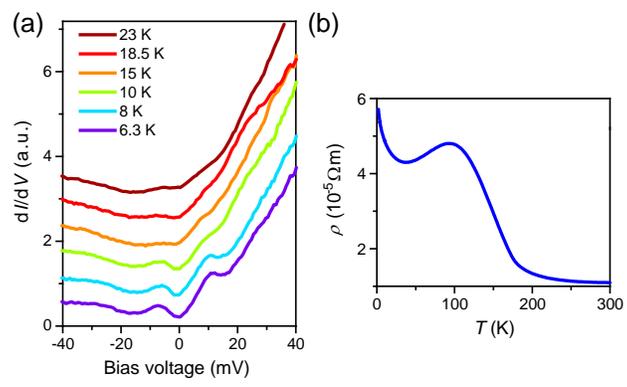

Figure 4

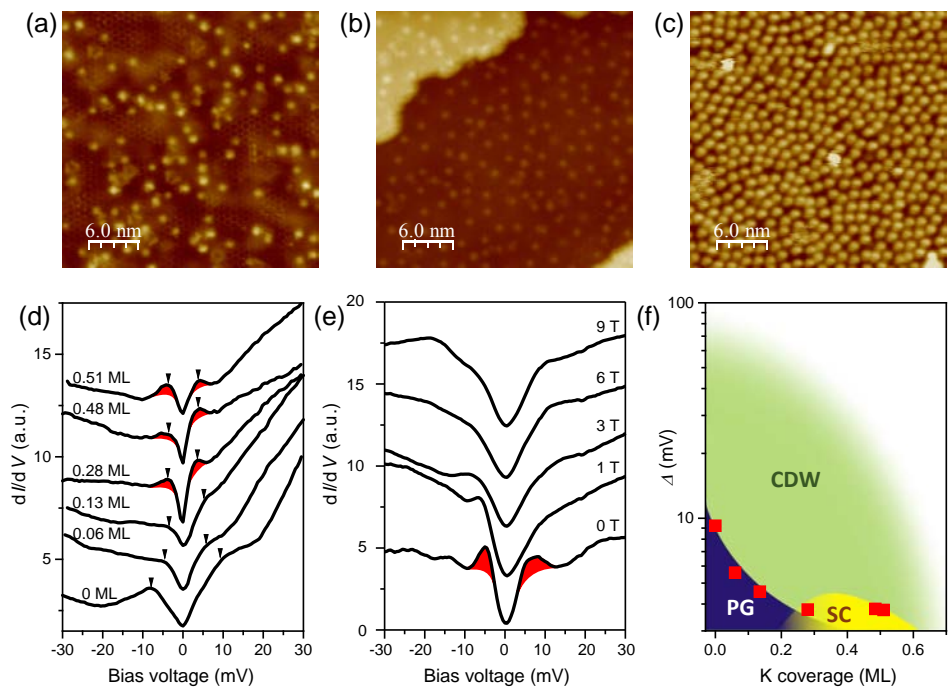

Figure 5